\documentclass[10pt,onecolumn,amsmath,amssymb,floatfix, notitlepage]{revtex4-1}
\usepackage{upgreek}
\usepackage[T1]{fontenc}
\usepackage[utf8]{inputenc}
\usepackage[right=0.75in, left=0.75in, top=1in]{geometry}
\usepackage{microtype,bm,bbm,graphicx,booktabs,times}
\usepackage[usenames,dvipsnames]{xcolor}

\usepackage{setspace}
\selectfont{}
\usepackage{hyperref}
\hypersetup{colorlinks,allcolors=Black,linktoc=all}

\usepackage[capitalize]{cleveref}
\crefformat{section}{Sec.~#2#1#3}

\newcommand{\ie}{{\it i.e.}}

\newcommand*\ket[1]{|#1\rangle}
\newcommand*\bra[1]{\langle #1|}

\newcommand{\edit}[1]{{\color{black} #1}}

\begin{document}
\title[\resizebox{3in}{!}{Long title}]{Analytic and geometric properties of scattering from periodically modulated quantum-optical systems}
\author{Rahul Trivedi}\email{rtrivedi@stanford.edu}
\affiliation{E. L. Ginzton Laboratory, Stanford University, Stanford, CA 94305, USA}
\author{Alex White}
\affiliation{E. L. Ginzton Laboratory, Stanford University, Stanford, CA 94305, USA}
\author{Shanhui Fan}
\affiliation{E. L. Ginzton Laboratory, Stanford University, Stanford, CA 94305, USA}
\author{Jelena Vu\v{c}kovi\'c}
\affiliation{E. L. Ginzton Laboratory, Stanford University, Stanford, CA 94305, USA}
\date{\today}
\begin{abstract}
    We study the scattering of photons from periodically modulated quantum-optical systems. For excitation-number conserving quantum optical systems, we connect the analytic structure of the frequency-domain $N$-photon scattering matrix of the system to the Floquet decomposition of its effective Hamiltonian. Furthermore, it is shown that the first order contribution to the transmission or equal-time $N-$photon correlation spectrum with respect to the modulation frequency is completely geometric in nature \ie~it only depends on the Hamiltonian trajectory and not on the precise nature of the modulation being applied.
\end{abstract}
\maketitle

\section{Introduction}
Quantum information processing and communication systems rely strongly on the generation and manipulation of non-classical states of light \cite{lodahl2015interfacing,ding2016demand,michler,senellart2017high,zhang2018strongly, kok2007linear,o2009photonic,roy2017colloquium,reiserer2015cavity,duan2010colloquium,sangouard2011quantum,nemoto2014photonic}. Implementing quantum systems for such applications often involves interfacing a localized quantum system (e.g.~a few-level system such as a quantum dot or color center) with bosonic baths (such as optical fibers or waveguides). Significant control over the states of light emitted by the localized system into the bosonic bath can be gained by engineering the coupling between the two \cite{englund2005controlling,daveau2017efficient, pelton2002efficient}, and by controlling the excitation of the localized system \cite{he2013demand, fischer2017signatures, hanschke2018quantum}. Recently, the ability to modulate the localized system on frequency-scales comparable to or exceeding the decay rate of the localized system into the bosonic bath has been demonstrated in various quantum-optical platforms such as quantum dots \cite{metcalfe2010resolved} and color centers \cite{miao2019electrically}. \edit{This has opened up the possibility of engineering the spectral content of the photons scattered by the localized system into the bosonic bath by engineering the modulation applied on the localized system. Such spectral engineering could enable quantum networks of localized systems that are robust to variations in their physical characteristics, unlock quantum information protocols relying on high-dimensional entangled photon states \cite{pichler2016photonic} and realize non-reciprocal photon transport \cite{yuan2015achieving}.

From a theoretical standpoint, it has opened up the question of how to calculate and understand the scattering properties of the modulated localized system.} The scattering properties of time-independent (unmodulated) localized systems can be completely described by its scattering matrix. Significant progress has been made in developing single and two-photon scattering matrices for specific localized systems (e.g.~two-level systems, Jaynes Cumming systems) by adapting a variety of different techniques from quantum field-theory \cite{shen2007strongly, shi2013two, shi2009lehmann, shi2015multiphoton}. The problem of systematically calculating scattering and emission from a general time-independent Markovian localized system was addressed in refs.~\cite{caneva2015quantum, xu2015input}, and it was shown that the computation of scattering matrices only required diagonalization of an effective non-hermitian Hamiltonian that is completely restricted to the Hilbert space of the localized system. The introduced formalism can be used to derive explicit relationships between the few-photon scattering properties of the localized system and the spectrum of its effective Hamiltonian \cite{xu2013analytic, trivedi2019photon} and this has been employed to understand a number of experimentally relevant quantum systems \cite{trivedi2019photon, xu2018generate}.

While most of the efforts in calculating and understanding scattering matrices were restricted to time-independent localized systems, a procedure for calculating the propagator from pulsed localized systems (i.e.~systems whose Hamiltonian has time-dependence only within a finite time-window) was recently developed \cite{trivedi2018few, fischer2017scattering}. It was shown that it is possible to define a scattering matrix for a time-dependent system provided it is asymptotically time-independent, and a recipe for its computation was provided \cite{trivedi2018few}. This procedure was applied to understand scattering of a single-photon from a two-level system driven by a pulsed laser, and the scattering matrix was shown to have significantly different structure from that of a time-independent two-level system \cite{trivedi2018few}.

In this paper, we consider the problem of calculating the scattering matrix for a periodically modulated localized system. We focus exclusively on localized systems which are excitation number conserving even in the presence of periodic modulation, and relate the scattering matrices to the Floquet decomposition of the non-Hermitian effective Hamiltonian of the localized system. Special attention is paid to the difference in the analytic properties of the resulting scattering matrix from the scattering matrix of time-independent systems. Finally, we consider the slow modulation regime and study the properties of the equal-time $N-$photon correlation function. It is shown that this correlation function, to the zeroth order in the modulation frequency, is equal to the time-average of the instantaneous correlation function obtained by assuming the system to be time independent and that the first order correction is completely geometric in nature.

This paper is organized into three major sections --- section \ref{sec:model} introduces the mathematical model of the system under consideration along with a review of the frequency-domain scattering matrix. Section \ref{sec:smat} presents the construction and general properties of the $N-$photon scattering matrix for a periodically modulated quantum system. As an example, single and two-photon scattering from a cavity with Kerr-nonlinearity is studied. Finally, in section \ref{sec:slow_mod_smat}, we study the properties of the equal-time $N-$photon correlation function in the slow modulation regime.

\section{Model and prelimnaries}\label{sec:model}
This section is intended to introduce the model for the system under consideration and also provide a review of scattering theory for open quantum systems. The analysis in this section closely follows that of refs.~\cite{xu2015input, trivedi2018few}.

\begin{figure}[b]
\centering
\includegraphics[scale=0.4]{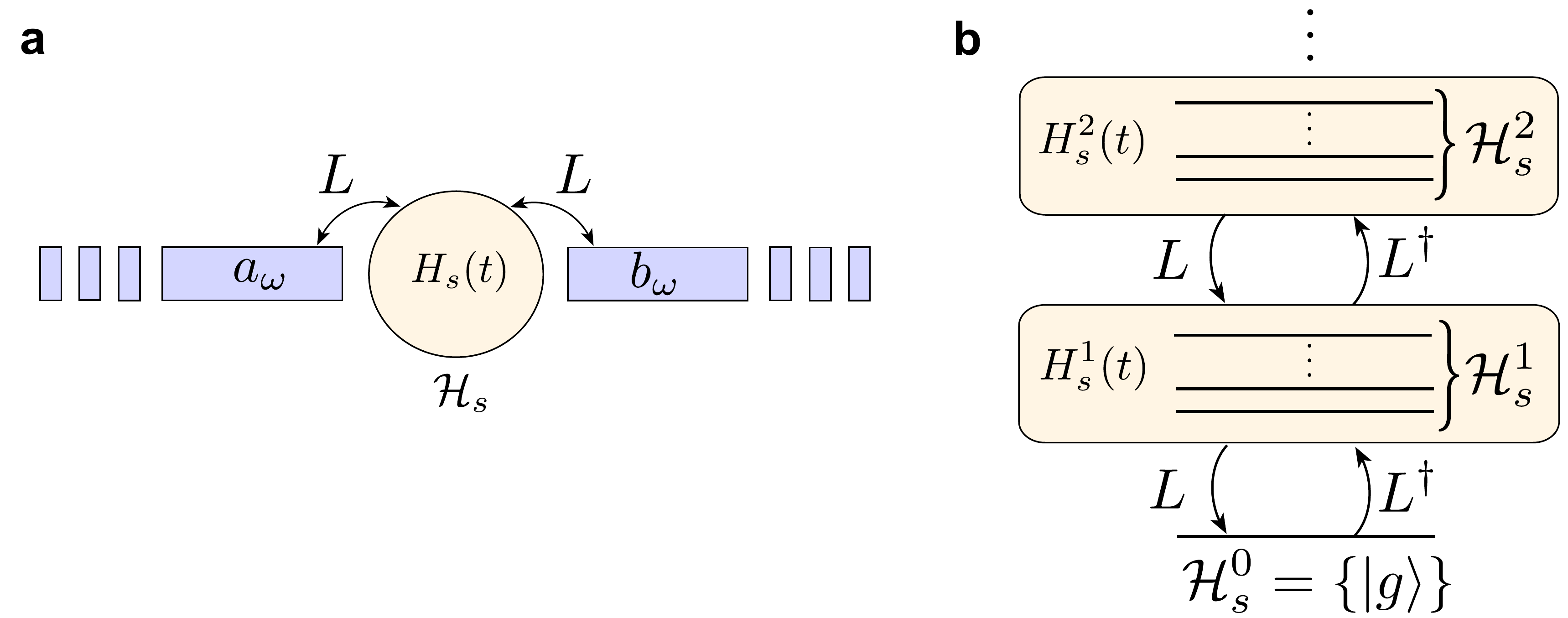}
\caption{\textbf{Schematic:} \textbf{a.} Schematic of the a modulated localized system coupling to the bosonic baths. The Hilbert space of the localized system is denoted by $\mathcal{H}_s$ and the Hilbert space of the bosonic baths are described by the frequency-dependent annihilation operators $a_\omega$ and $b_\omega$. The Hamiltonian of the localized system is denoted by $H_s(t)$, and $L$ is the system operator through which the localized system couples bosonic baths. b) Schematic of the level-structure of an excitation-number conserving localized system. The Hilbert space of the system can be divided into $\mathcal{H}_s^0, \mathcal{H}_s^1, \mathcal{H}_s^2\dots$, which correspond to space of states with excitation numbers 0, 1, 2 $\dots$. The operator $L$ maps $\mathcal{H}_s^n$ to $\mathcal{H}_s^{n-1}$ for $n\geq 1$ and annihilate the states in $\mathcal{H}_s^0$ while the operators $L^\dagger$ maps $\mathcal{H}_s^n$ to $\mathcal{H}_s^{n + 1}$.}
\label{fig:schematic}
\end{figure}

We consider a general class of time-dependent systems which have a periodic localized system interacting with two bosonic baths schematically depicted in Fig.~\ref{fig:schematic}a. The Hilbert space of the two bosonic baths is described by frequency-dependent annihilation operators $a_\omega$ and $b_\omega$. These operators satisfy the bosonic commutation relations: $[a_\omega, a_\nu] = 0, [b_\omega, b_\nu] = 0, [a_\omega, a_\nu^\dagger] = \delta(\omega - \nu), [b_\omega, b_\nu^\dagger] = \delta(\omega - \nu)$ and $[a_\omega, b_\nu] = [a_\omega, b_\nu^\dagger] = 0$. The dynamics of this system are governed by the following time-dependent Hamiltonian:
\begin{align}\label{eq:basic_hamiltonian}
H(t) = H_s(t) + \sum_{s \in \{a, b\}} \int_{-\infty}^\infty \omega s_\omega^\dagger s_\omega d\omega + \sum_{s\in \{a, b\}} \int \textrm{i}\big(L s_\omega^\dagger - s_\omega L^\dagger \big)\frac{d\omega}{\sqrt{2\pi}}
\end{align}
Here $L$ is the operator through which the localized system couples to the bosonic baths. Throughout this paper, we will consider the bath described by $a_\omega$ as the input bath and that described by $b_\omega$ as the output bath. For simplicity, we assume that the two baths couple equally to the localized system.

We will also assume the localized system and its coupling to the bosonic baths to be excitation number conserving --- this requires that (a) the Hilbert space $\mathcal{H}_s$ of the localized system can be expressed as a direct sum of subspaces: $\mathcal{H}_s = \mathcal{H}_s^0 \oplus \mathcal{H}_s^1 \oplus \mathcal{H}_s^2\dots $ such that each subspace $\mathcal{H}_s^n$ is invariant under evolution with respect to the system Hamiltonian $H_s(t)$ and (b) $L$ maps the subspace $\mathcal{H}_s^n$ to $\mathcal{H}_s^{n-1}$ for $n \geq 1$, $L^\dagger$ maps the subspace $\mathcal{H}_s^{n}$ to $\mathcal{H}_s^{n+1}$ for $n \geq 0$ and $L$ annihilates $\mathcal{H}_s^0$ \ie~$\mathcal{H}_s^0$ is within the null-space of $L$. Throughout this paper, we will refer to $\mathcal{H}_s^n$ as the $n^\text{th}$ excitation subspace and associate with it an excitation number $n$. The operator $L$ then decreases the excitation number of the localized system's state by 1 and $L^\dagger$ increases it by 1. Furthermore, evolving a state in $\mathcal{H}_s^0$ with respect to $H(t)$ is identical to evolving it with respect to $H_s(t)$ without interacting with the bosonic bath --- $\mathcal{H}_s^0$ is therefore the space of the ground states of the localized system. In this paper, we will restrict ourselves to systems with a single ground state $\ket{g}$ \ie~$\mathcal{H}_s^0 = \{\ket{g}\}$.

As is shown in appendix \ref{app:ph_no_cons}, for an excitation-number conserving system with a single ground state, $N$ photons incident on the localized system can only scatter into $N$ outgoing photons, and consequently the scattering properties of the system can be described by the $N$-photon scattering matrix:
\begin{align}\label{eq:smat_elem}
S(\omega_1, \omega_2 \dots \omega_N; \nu_1, \nu_2 \dots \nu_N) = \bra{\text{vac}; g}\bigg( \prod_{i=1}^N b_{\omega_i} \bigg)\hat{\textrm{S}} \bigg(\prod_{i=1}^N a_{\nu_i}^\dagger \bigg)\ket{\text{vac}; g}
\end{align}
where $\hat{\textrm{S}}$ is the scattering-matrix defined via \cite{taylor2006scattering}:
\begin{align}\label{eq:smat_def}
\hat{\textrm{S}} = \lim_{\substack{t_+ \to \infty \\ t_- \to -\infty}} U_0(t_0, t_+) U(t_+, t_-) U_0(t_-, t_0)
\end{align}
Here $U(\cdot, \cdot)$ is the propagator corresponding to the Hamiltonian $H(t)$ and $U_0(\cdot, \cdot)$ is the propagator corresponding to the Hamiltonian $H_0(t)$ corresponding to the uncoupled localized system and bosonic baths:
\begin{align}\label{eq:ref_hamil}
H_0(t) = H_s(t) + \sum_{s\in \{a, b\}} \int_{-\infty}^\infty \omega s_\omega^\dagger s_\omega d\omega
\end{align}
Additionally, $t_0$ is a time-reference that is used for defining the input and output asymptotes corresponding to the states incident and scattered from the localized system \cite{taylor2006scattering}. While this time reference does not affect the scattering matrix of a time-independent system, it encodes the `time of arrival' of the incident photon wave-packet and thus is relevant for time-dependent system. Using the input-output formalism \cite{gardiner1985input, xu2015input}, it can easily be shown that the scattering matrix element in Eq.~\ref{eq:smat_elem} is related to the Heisenberg-picture system operator $L(t) = U(t_0, t) L\ U(t, t_0)$ via (refer to appendix \ref{app:smat_to_gfunc} for derivation):
\begin{align}\label{eq:gfunc_to_smat}
S(\omega_1, \omega_2 \dots \omega_N; \nu_1, \nu_2 \dots \nu_N) =(-1)^N e^{-\textrm{i}\phi(t_0)} \int_{-\infty}^\infty \dots \int_{-\infty}^\infty G(t_1, t_2 \dots t_N; s_1, s_2 \dots s_N) \prod_{i=1}^N e^{\textrm{i}(\omega_i t_i - \nu_i s_i)}\frac{dt_i ds_i}{2\pi}
\end{align}
where $\phi(t_0) = \sum_{i=1}^N(\omega_i - \nu_i)t_0 $ and we have introduced the time-domain system Green's function:
\begin{align}\label{eq:gfunc}
G(t_1, t_2 \dots t_N; s_1, s_2 \dots s_N) = \bra{\text{vac}; g} \mathcal{T} \bigg[\prod_{i=1}^N L(t_i) \prod_{i=1}^N L^\dagger(s_i) \bigg] \ket{\text{vac}; g}
\end{align}
where $\mathcal{T}[\cdot]$ indicates time-ordering in its arguments. An application of the quantum regression theorem can allow us to evaluate these Green's functions entirely within the Hilbert space of the localized system by replacing the Heisenberg operators $L(t)$ and $L^\dagger(t)$ with respect to the Hamiltonian $H(t)$ with Heisenberg operators $\tilde{L}(t)$ and $\tilde{L^\dagger}(t)$ with respect to the non-Hermitian effective Hamiltonian $H_\text{eff}(t) = H_s(t) - \textrm{i}L^\dagger L $ \ie
\begin{subequations}\label{eq:gfunc_eff_hamil}
\begin{align}
G(t_1, t_2 \dots t_N; s_1, s_2 \dots s_N) = \bra{g} \mathcal{T}\bigg[\prod_{i=1}^N \tilde{L}(t_i) \prod_{i=1}^N \tilde{L^\dagger}(s_i) \bigg] \ket{g},
\end{align}
where
\begin{align}
\tilde{O}(t) = U_\text{eff}(0, t) O U_\text{eff}(t, 0)\ \text{for} \ O \in \{L, L^\dagger\},
\end{align}
\end{subequations}
with $U_\text{eff}(t, s)$ being the propagator corresponding to $H_\text{eff}(t)$. It can be noted that both $H_\text{eff}(t)$ and $U_\text{eff}(t, s)$ do not effect the excitation number of the state that they act on, and can therefore be described by their restrictions $H_\text{eff}^n(t)$ and $U_\text{eff}^n(t, s)$ respectively within the $n^\text{th}$ excitation subspace $\mathcal{H}_s^n$. Eqs.~\ref{eq:gfunc_to_smat} and \ref{eq:gfunc_eff_hamil} are used in the following sections to study the computation and properties of the frequency-domain scattering matrix.

\section{Scattering matrices}\label{sec:smat}
In this section, we explore the systematic construction of the frequency-domain scattering matrices (Eq.~\ref{eq:gfunc_to_smat}) of the modulated quantum system. Special attention is paid to the similarities and differences that arise in these scattering matrices relative to the time-independent case. Explicit results are provided for single and two-photon scattering matrices.

\subsection{Construction and analytic properties}
The discrete time-translation symmetry of the periodically modulated quantum system imposes a fundamental constraint on the structure of the $N-$photon frequency-domain scattering matrix. In particular, the form of the scattering matrix should conserve the total photon frequency modulo $\Omega$ --- this implies that the scattering matrix $S(\omega_1, \omega_2 \dots \omega_N; \nu_1, \nu_2 \dots \nu_N)$ can be written as a sum of terms proportional to $\delta(\sum_{i=1}^N \omega_i - \sum_{i=1}^N \nu_i  - k\Omega)$:
\begin{align}\label{eq:smat_general_decomp}
S(\omega_1, \omega_2 \dots \omega_N; \nu_1, \nu_2 \dots \nu_N) = \sum_{k=-\infty}^{\infty} e^{-\textrm{i}k\Omega t_0} S_k(\omega_1, \omega_2 \dots \omega_N; \nu_1, \nu_2 \dots \nu_N)\delta\bigg(\sum_{i=1}^N \omega_i - \sum_{i=1}^N \nu_i  - k\Omega\bigg).
\end{align}
Note that here we have explicitly shown the dependence on the time-reference $t_0$ used for defining the scattering matrix (Eq.~\ref{eq:smat_def}) that enters Eq.~\ref{eq:gfunc_to_smat} as a phase factor depending on the difference between the total input and output frequencies under consideration --- the discrete time-translation symmetry of the Hamiltonian ensures that the scattering matrix is periodic in $t_0$ with period $2\pi / \Omega$. This general form of the scattering matrix can be contrasted with the scattering matrix for time-independent systems, which would be proportional to $\delta(\sum_{i=1}^N \omega_i - \sum_{i=1}^N \nu_i)$ since it conserves the total photon frequency and consequently be independent of the time-reference $t_0$.

The functions $S_k(\omega_1, \omega_2 \dots \omega_N; \nu_1, \nu_2 \dots \nu_N)$ in Eq.~\ref{eq:smat_general_decomp} can, in general, be further decomposed into a sum of a non-singular function, denoted by $S_k^\text{C}(\omega_1, \omega_2 \dots \omega_N; \nu_1, \nu_2 \dots \nu_N)$, and terms with delta-function singularities. The \emph{connected part} of the $N-$photon scattering matrix, $S^\text{C}(\omega_1, \omega_2 \dots \omega_N; \nu_1, \nu_2 \dots \nu_N)$, can then be defined as:
\begin{align}
S^\text{C}(\omega_1, \omega_2 \dots \omega_N; \nu_1, \nu_2 \dots \nu_N) = \sum_{k=-\infty}^\infty e^{-\textrm{i}k\Omega t_0}  S_k^\text{C}(\omega_1, \omega_2 \dots \omega_N; \nu_1, \nu_2 \dots \nu_N)\delta\bigg(\sum_{i=1}^N \omega_i - \sum_{i=1}^N \nu_i  - k\Omega\bigg).
\end{align}
From a physical standpoint,  $S^\text{C}(\omega_1, \omega_2 \dots \omega_N; \nu_1, \nu_2 \dots \nu_N)$ accounts for all the nonlinear interactions between the $N$ incident photons that are induced by the localized quantum system --- while it conserves the total photon frequency modulo $\Omega$, it can in general lead to a change in the individual photon frequencies. Furthermore, an application of the cluster decomposition principle allows us to construct the full $N-$photon scattering matrix from its connected part and the connected part of fewer photon scattering matrices \cite{xu2013analytic, xu2015input}:
\begin{align}\label{eq:smat_cluster_decomp}
S(\omega_1, \omega_2 \dots \omega_N; \nu_1, \nu_2 \dots \nu_N) = \sum_{\mathcal{B}}\sum_P \prod_{k=1}^{|\mathcal{B}|} S^\text{C}(\omega_{\mathcal{B}_kP(1)}, \omega_{\mathcal{B}_kP(2)} \dots; \nu_{\mathcal{B}P_k(1)}, \nu_{\mathcal{B}P_k(2)} \dots ),
\end{align}
where $\mathcal{B}$ is an ordered partition of $\{1, 2, 3\dots N\}$ into smaller subsets, $P$ is a permutation of $\{1, 2, 3\dots \}$ and $\mathcal{B}{P}$ is the partition $\mathcal{B}$ applied on $\{P(1), P(2) \dots P(N)\}$.

For time-independent localized system, it can be shown that the frequency domain scattering matrix is completely determined by the spectral decomposition of the effective Hamiltonian of the localized system. In particular, the position and linewidth of resonances in the $N-$photon scattering matrix are determined by the complex eigenvalues of the effective Hamiltonian, and the amplitude of the scattering matrix at these resonances is determined by its eigenvectors. For periodically modulated localized systems, a similar relationship can be established between the scattering matrices and the Floquet decomposition of the effective Hamiltonian. Since the effective Hamiltonian is non-Hermitian, its Floquet decomposition within the $n-$excitation subspace requires the solution of the following eigenvalue equations \cite{longhi2017floquet}:
\begin{subequations}\label{eq:floquet_problem}
\begin{align}
&H_\text{eff}^n(t)\ket{\phi_k^n(t)} + \textrm{i}\frac{d}{dt} \ket{\phi_k^n( t)} = \lambda_k^n \ket{\phi_k^n(t)} \\
&\big(H_\text{eff}^n(t)\big)^\dagger\ket{\chi_k^n(t)} - \textrm{i}\frac{d}{dt} \ket{\chi_k^n(t)} = \big(\lambda_k^n\big)^* \ket{\chi_k^n(t)}
\end{align}
\end{subequations}
where $\lambda_k^n$ is the $k^\text{th}$ Floquet eigenvalue of $H_\text{eff}^n(t)$ and $(\ket{\phi_k^n(t)}, \ket{\chi_k^n(t)})$ are the $k^\text{th}$ birothogonal Floquet eigenvectors of $H_\text{eff}^n$. We note that both $\ket{\phi_k^n(t)}$ and $\ket{\chi_k^n(t)}$ are periodic with periodicity of the system Hamiltonian: $\ket{\phi_k^n(t + T)} = \ket{\phi_k^n(t)}$ and $\ket{\chi_k^n(t + T)} = \ket{\chi_k^n(t)}$. They also satisfy $\bra{\chi_k^n(t)}\phi_l^n(t)\rangle = \delta_{k, l}$ for all $t \in (0, T]$. The Floquet eigenvalue $\lambda_k^n$ will be, in general, a complex number and can be expressed in terms of its real and imaginary parts: $\lambda_k^n = \varepsilon_k^n - \textrm{i}\kappa_k^n / 2$. We note that $\varepsilon_k^n$ can only be uniquely specified to modulo $\Omega$. Provided such biorthogonal states exist, the propogator $U_\text{eff}^n(t, s)$ in the $n^\text{th}$ excitation subspace can be expressed as:
\begin{align}\label{eq:floquet_decomp}
U_\text{eff}^n(t, s) = \sum_k \ket{\phi_k^n(t)} \bra{\chi_k^n(s)} \exp(-\textrm{i}\lambda_k^n(t - s)).
\end{align}
This decomposition of the propagator along with Eq.~\ref{eq:gfunc_eff_hamil} can be used to relate the frequency domain scattering matrices to the Floquet decomposition of the effective Hamiltonian. Due to the periodic time-dependence of the Floquet states, the frequency-domain scattering matrices has resonances at $\varepsilon_k^n + p\Omega$ for $p \in \mathbb{Z}$ with linewidths $\kappa_k^n$. Furthermore, the amplitudes of these resonances are determined by the Fourier components of the periodic Floquet eigenstates $\ket{\phi_k^n(t)}$ and $\ket{\chi_k^n(t)}$. This is made more explicit for single and two-photon scattering matrices in the following subsection.

\subsection{Single and two-photon scattering matrices}

Of particular interest are the single and two-photon scattering matrices, since they can often be easily probed experimentally with transmission and two-photon correlation experiments. Following the procedure outlined above for the single-photon scattering matrix, we obtain (details in appendix \ref{app:smat}):
\begin{subequations}\label{eq:conn_part_one_ph}
\begin{align}
S(\omega; \nu) = \sum_{k\in \mathbb{Z}} e^{-\textrm{i}k\Omega t_0} S_k(\nu) \delta(\omega - \nu - k\Omega),
\end{align}
with
\begin{align}
S_k(\nu) =  \sum_{m\in \mathbb{Z}}\text{L}^{1\to 0}_{k+m}  \text{D}\bigg(\frac{1}{\textrm{i}(\upvarepsilon^1 + m\Omega - \nu) + \upkappa^1 / 2} \bigg) \text{L}^{0\to 1}_m.
\end{align}
\end{subequations}
Here $\upvarepsilon^1$ and $\upkappa^1$ are vectors of $\varepsilon_n^1$ and $\kappa_n^1$ respectively and $\text{D}(\cdot)$ constructs a diagonal matrix from a vector that is passed as its argument. $\text{L}^{1\to0}_k$ is a row vector and $\text{L}^{0\to 1}_k$ is a column vector, and their elements are given by:
\begin{align}\label{eq:L_fc_def_single_ex}
\big[\text{L}^{1\to 0}_k\big]_n = \int_0^T \bra{g} L \ket{\phi_n^1(t)} e^{\textrm{i}k\Omega t} \frac{dt}{T} \ \text{and} \ \big[\text{L}^{0\to 1}_k\big]_n = \int_0^T \bra{\chi_n^1(t)} L^\dagger \ket{g}e^{-\textrm{i}k\Omega t}\frac{dt}{T}
\end{align}
Clearly, the form of the single-photon scattering matrix implies that a photon at frequency $\nu$ is in general scattered into photons at frequencies differing from $\nu$ by an integer multiple of $\Omega$. Furthermore, the amplitude of transmission at these sidebands would in general depend on the Fourier series components of the Floquet states $\ket{\phi_n^1(t)}$ and $\ket{\chi_n^1(t)}$.

A similar procedure can be followed for the computation of the two-photon scattering matrix. As is shown in appendix \ref{app:smat}, the connected part of the two-photon scattering matrix, $S^\text{C}(\omega_1, \omega_2; \nu_1, \nu_2)$ can be expressed as a sum of two components: $S^{\text{C}, 1}(\omega_1, \omega_2; \nu_1, \nu_2)$ which is completely determined by the Floquet-decomposition of $H_\text{eff}^1(t)$ and $S^{\text{C}, 2}(\omega_1, \omega_2; \nu_1, \nu_2)$ which depends on the Floquet-decomposition of $H_\text{eff}^2(t)$:
\begin{subequations}\label{eq:conn_part_two_ph}
\begin{alignat}{2}
S^{\text{C}, 1}_k(\omega_1, \omega_2; \nu_1, \nu_2) &= \frac{1}{2\pi \textrm{i}}\sum_{P, Q}\sum_{p, m, n \in \mathbb{Z}} \bigg[&&\text{L}^{1\to 0}_p \text{D}\bigg(\frac{1}{\textrm{i}(\upvarepsilon^1 - \omega_{P(1)} + p\Omega) + \upkappa^1/2} \bigg)\text{L}^{0\to 1}_{n + p - k} \mathcal{P}\frac{1}{\omega_{P(2)} - \nu_{Q(2)}-n\Omega} \nonumber \\& &&\text{L}^{1\to 0}_{m + n} \text{D}\bigg(\frac{1}{\textrm{i}(\upvarepsilon^1 - \nu_{Q(2)} + m\Omega) + \upkappa^1 / 2} \bigg)\text{L}^{0\to 1}_{m}\bigg],\\
S^{\text{C}, 2}_k(\omega_1, \omega_2; \nu_1, \nu_2) &= \frac{1}{2\pi}\sum_{P, Q}\sum_{p, m, n \in \mathbb{Z}} \bigg[&&\text{L}^{1\to 0}_p \text{D}\bigg(\frac{1}{\textrm{i}(\upvarepsilon^1 - \omega_{P(1)} + p\Omega) + \upkappa^1/2} \bigg)\text{L}^{2\to 1}_{n - p + k} \text{D}\bigg(\frac{1}{\textrm{i}(\upvarepsilon^2 - \nu_1 - \nu_2 + n\Omega) + \upkappa^2/2} \bigg) \nonumber \\& &&\text{L}^{1\to 2}_{n - m} \text{D}\bigg(\frac{1}{\textrm{i}(\upvarepsilon^1 - \nu_{Q(2)} + m\Omega) + \upkappa^1 / 2} \bigg)\text{L}^{0\to 1}_{m}\bigg],
\end{alignat}
\end{subequations}
where $P, Q$ are permutations of the two-element set $\{1, 2\}$, $\mathcal{P}$ indicates the principal part, $\upvarepsilon^2$ and $\upkappa^2$ are vectors of $\varepsilon_n^2$ and $\kappa_n^2$ and $\textrm{L}^{2\to 1}_n, \textrm{L}^{1\to2}_n$ are matrices whose elements are given by:
\begin{align}\label{eq:L_fc_def_two_ex}
\big[\text{L}^{2\to 1}_k]_{m, n} = \int_0^T \bra{\chi_m^1(t)} L \ket{\phi_n^2(t)} e^{\textrm{i}k\Omega t} \frac{dt}{T} \ \text{and} \ \big[\text{L}^{1\to 2}_k]_{m, n} = \int_0^T \bra{\chi_m^2(t)}L^\dagger \ket{\phi_n^1(t)}e^{-\textrm{i}k\Omega t}\frac{dt}{T}
\end{align}
The full two-photon scattering matrix can be constructed from the connected parts in Eqs.~\ref{eq:conn_part_one_ph} and \ref{eq:conn_part_two_ph} by an application of the cluster decomposition principle (Eq.~\ref{eq:smat_cluster_decomp}). We note that while it appears that $S^{\text{C}, 1}(\omega_1, \omega_2; \nu_1, \nu_2)$ has singularites corresponding to principle parts --- as is shown in Appendix \ref{app:smat}, a proper evaluation of the summation removes these singularities.

As an illustrative example of this procedure, we consider the computation of the single- and two-photon scattering matrices for a cavity with Kerr-nonlinearity and a periodically modulated resonance frequency. The Hamiltonian of the localized system under consideration here is given by:
\begin{align}
H_s(t) =  \Delta(t)a^\dagger a + \chi(a^\dagger)^2 a^2,
\end{align}
with a coupling operator $L = \sqrt{\kappa / 2} \ a$. Here, $\Delta(t)$ is the periodic modulation applied on the cavity mode, $\chi$ is the photon-photon repulsion in the cavity due to the Kerr nonlinearity and $\kappa$ is the decay rate of the cavity. We assume that the mean of $\Delta(t)$ over one period is 0. Since the $N-$excitation subspace for this system has dimensionality $1$, there is only one solution to the Floquet problem in Eq.~\ref{eq:floquet_problem}:
\begin{align}
\ket{\phi^N_1(t)} = \ket{\chi^N_1(t)} = e^{-\textrm{i}N \varphi(t)} \frac{(a^\dagger)^N}{\sqrt{N!}}\ket{g} \ \text{and} \ \varepsilon^N_1 = - \frac{\textrm{i}N\kappa}{2} + \chi N (N - 1),
\end{align}
where $\varphi(t) = \int_0^t \Delta(t')dt'$. With this choice of Floquet states, $\text{L}^{1\to 0}_k, \text{L}^{0\to 1}_k, \text{L}^{2 \to 1}_k$ and $\text{L}^{1 \to 2}_k$ in Eqs.~\ref{eq:L_fc_def_single_ex} and \ref{eq:L_fc_def_two_ex} reduce to scalars given by:
\begin{align}
\text{L}^{1\to 0}_k = \sqrt{\kappa}\alpha_k, \ \text{L}^{2\to 1}_k = \sqrt{2\kappa}\alpha_k, \ \text{L}^{1\to 2}_k = \sqrt{2\kappa}\alpha_k^*, \text{ and } \text{L}^{0\to 1}_k = \sqrt{\kappa}\alpha_k^*
\end{align} 
where $\alpha_k$ are the Fourier-series components of $e^{-\textrm{i}\varphi(t)}$:
\begin{align}
e^{-\textrm{i}\varphi(t)} = \sum_{k\in \mathbb{Z}} \alpha_k e^{-\textrm{i}k\Omega t}.
\end{align}
Therefore, the single-photon scattering matrix (Eq.~\ref{eq:conn_part_one_ph}) evaluates to
\begin{align}
S_k(\nu) = \sum_{m\in\mathbb{Z}} \frac{\alpha_{k + m} \alpha_m^*}{\textrm{i}( m\Omega - \nu) + \kappa / 2}.
\end{align}
Similarly, using Eqs.~\ref{eq:conn_part_two_ph}, the two-photon scattering matrices evaluate to:
\begin{subequations}
\begin{alignat}{2}
S^{\text{C}, 1}_k(\omega_1, \omega_2; \nu_1, \nu_2) &= \frac{\kappa^2}{2\pi \textrm{i}}\sum_{P, Q}\sum_{p, m, n \in \mathbb{Z}} \bigg[ \frac{\alpha_p \alpha^*_{n + p -k}\alpha_{m + n}\alpha^*_m}{\big(\textrm{i}(p\Omega - \omega_{P(1)}) + \kappa / 2\big)\big(\textrm{i}(m\Omega - \nu_{Q(2)}) + \kappa/2 \big)}  \mathcal{P}\frac{1}{\omega_{P(2)} - \nu_{Q(2)}-n\Omega}  \bigg],\\
S^{\text{C}, 2}_k(\omega_1, \omega_2; \nu_1, \nu_2) &= \frac{\kappa^2}{\pi}\sum_{P, Q}\sum_{p, m, n \in \mathbb{Z}} \bigg[\frac{\alpha_p \alpha_{m-p+k}\alpha^*_{n-m}\alpha^*_m}{\big(\textrm{i}(p\Omega- \omega_{P(1)}) + \kappa/2\big)\big(\textrm{i}(n\Omega + 2\chi - \nu_1 - \nu_2) + \kappa\big)\big(\textrm{i}(m\Omega- \nu_{Q(2)} ) + \kappa/2\big)}\bigg],
\end{alignat}

Numerical studies of the single and two-photon transport through a modulated Kerr cavity with $\Delta(t) = \Delta_0 \sin \Omega t$ are shown  Figure \ref{fig:chi3_cavity_scat}. Figure~\ref{fig:chi3_cavity_scat}\textbf{a} shows the total single-photon transmission $T(\nu) = \sum_{k\in \mathbb{Z}} |S_k(\nu)|^2$ through the cavity for slow modulation ($\Omega \ll \kappa$) and fast modulation ($\Omega \gg \kappa$) of its resonant frequency. In the fast modulation regime, the transmission spectra shows resonances at integer multiples of $\Omega$ with the transmission being smaller than the resonant transmission for unmodulated cavity. In the slow modulation regime, moderate transmissions are achieved if the input photon is within the resonant frequencies achieved by the periodic modulation ($[-\Delta_0, \Delta_0]$). The amplitude $|S_k(\nu)|$ of a photon at frequency $\nu$ scattering into a photon at frequency $\nu + k\Omega$ within the slow and fast modulation regime is shown in Figs.~\ref{fig:chi3_cavity_scat}\textbf{b}. We point out that the single-photon transmissions obtained here are unaffected by the non-linearity $\chi$ in the cavity mode --- they are identical to the classical transmission that would be obtained through a linear cavity with the same modulation \cite{minkov2017exact}.
\begin{figure}[t]
\centering
\includegraphics[scale=0.28]{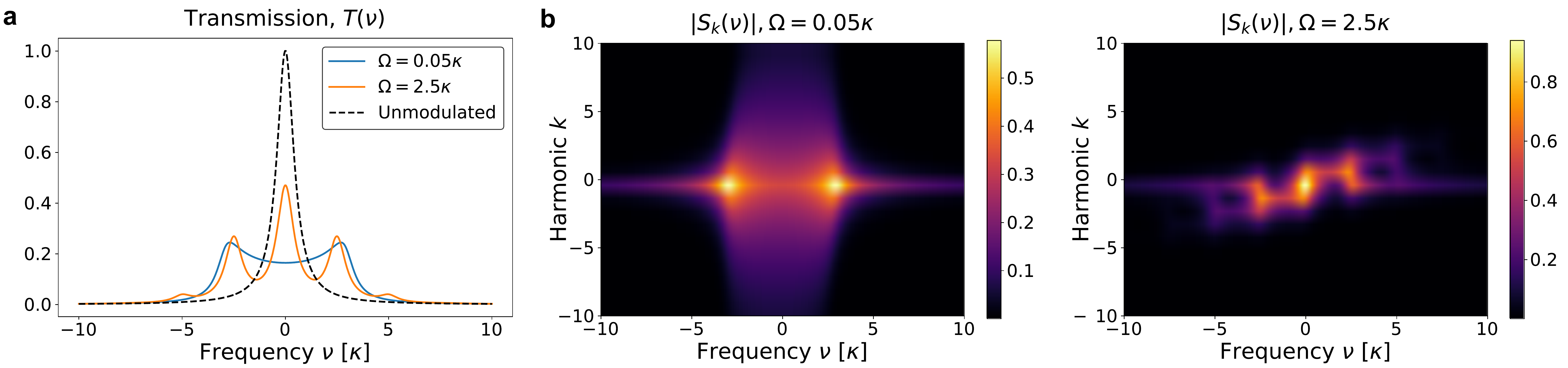}
\caption{\textbf{Single photon scattering from a modulated Kerr cavity.} \textbf{a.} The total single-photon transmission as a function of the input frequency $\nu$ for different modulation frequencies. \textbf{b.} Amplitude of scattering a single-photon at frequency $\nu$ into an output photon at  frequency $\nu + k\Omega$ as a function of $\nu$ and $k$. $\Delta_0 = 3\kappa$ has been assumed in all simulations.}
\label{fig:chi3_cavity_scat}
\end{figure}

The connected part of the two-photon scattering matrix corresponding to the $k^\text{th}$ side-band under excitation by two photons at frequencies $\nu_1 = \nu_2 = 0$, $S_k^\text{C}(\omega_1, \omega_2; \nu_1 = 0, \nu_2 = 0)$, is shown in Fig.~\ref{fig:chi3_cavity_scat_two_ph}. Since the output frequencies $\omega_1$ and $\omega_2$ of the two photons emitted into this sideband are constrained to satisfy $\omega_1 + \omega_2 = k\Omega$, they can be completely parametrized by their frequency different $\delta = \omega_1 - \omega_2$. As can be seen from Fig.~\ref{fig:chi3_cavity_scat_two_ph} --- the amplitude of the connected part increased on increasing the nonlinearity $\chi$. This is intuitively expected since the connected part captures the photon-photon interactions induced by the localized system. Furthermore, we note that there is an asymmetry in the amplitudes of the connected part corresponding to $k = 1$ and $k = -1$ --- this can be attributed to the fact that the nonlinearity $\chi$ results in an \emph{increase} in the cavity resonant frequency with the number of photons in the cavity and thus has larger contribution to one side-band as opposed to the other. Indeed, in the two-level system limit ($\chi\to \infty$), it can be seen from Fig.~\ref{fig:chi3_cavity_scat_two_ph} that both the sidebands have identical connected part amplitudes.
\begin{figure}[htpb]
\centering
\includegraphics[scale=0.28]{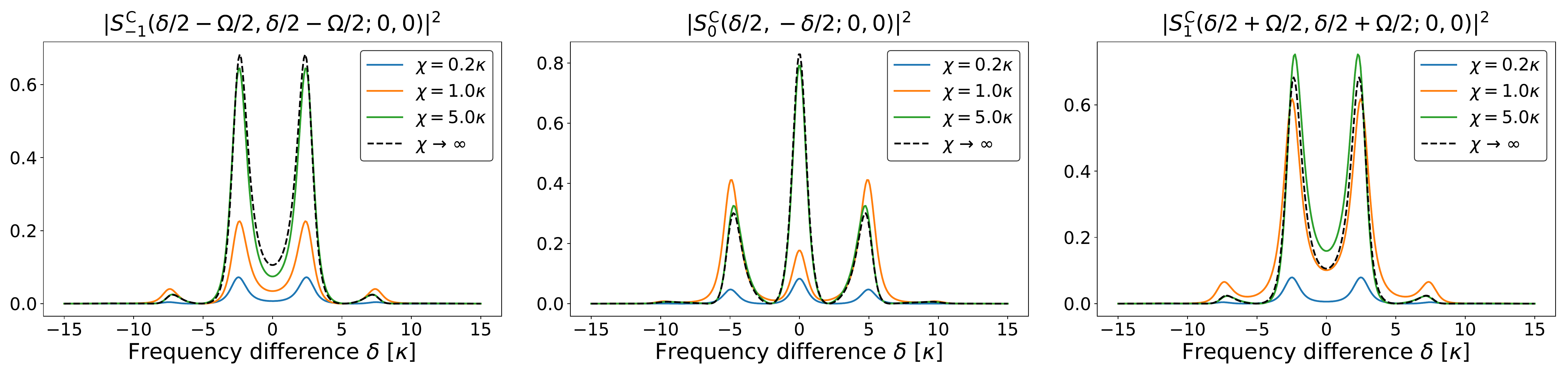}
\caption{\textbf{Two photon scattering from cavity with Kerr non-linearity:} The connected part of the two-photon scattering matrix for the $k = -1, 0, 1$ sidebands with the two input photons being at $\nu_1 = \nu_2 = 0$ as a function of the frequency difference $\delta$ between the output photons. Note that the two output photon scattered into the $k^\text{th}$ sideband are constrained have a mean frequency of $k{\Omega} / 2$. Parameter values of $\Omega= 2.5\kappa$ and $\Delta_0 = 3\kappa$ have been assumed in all simulations.}
\label{fig:chi3_cavity_scat_two_ph}
\end{figure}

\end{subequations}

\section{Geometric properties in slow modulation regime} \label{sec:slow_mod_smat}
In a number of physical systems, the modulation period is significantly smaller than the timescale of evolution of the localized system \cite{xiao2005berry, xiao2006berry}. Such systems are considered to be in the slow modulation regime and have been a subject of significant theoretical interest. In particular, for closed quantum systems, observables such as the Berry phase \cite{xiao2010berry} can be defined which only depend on the geometry of the Hamiltonian being modulated and are independent of the modulation being applied to the Hamiltonian. In this section, we study scattering from a slowly modulated quantum system. In particular, it is shown that the equal time $N-$photon correlation function, to the zeroth order in modulation frequency, is equal to the time-average of the instantaneous correlation function obtained by assuming the system to be time independent. Furthermore, we show that the first-order correction to the $N-$photon correlation function is purely geometric in nature \ie~it is independent of the precise form of the modulation applied on the Hamiltonian.

We consider a localized system with Hamiltonian dependent on a set of parameters $\text{p} = \{p_1, p_2 \dots p_M\}$: $H_s(\text{p})$. These parameters are varied along a closed loop $\mathcal{C}$ within the space of allowed parameters periodically to yield a Hamiltonian $H_s(t) = H_s(\text{p}(t))$. The equal-time $N-$photon correlation $G_N(\nu)$ at frequency $\nu$ is defined in terms of the $N-$photon scattering matrix via:
\begin{align}
G_N(\nu) = \frac{1}{N! T}\int_0^T \bigg|\int_{-\infty}^\infty \dots \int_{-\infty}^\infty S(\omega_1, \omega_2 \dots \omega_N; \nu, \nu \dots \nu)\prod_{i=1}^N e^{-\textrm{i}\omega_i t} d\omega_i \bigg|^2 dt,
\end{align}
or equivalently in terms of the $N-$excitation Green's function via:
\begin{align}\label{eq:nph_corr_gfunc}
G_N(\nu) = \frac{1}{N! T}\int_0^T \bigg| \int_{-\infty}^\infty \dots \int_{-\infty}^\infty G(t, t \dots t; s_1, s_2 \dots s_N) \prod_{i=1}^N e^{-\textrm{i}\nu s_i}ds_i \bigg|^2 dt.
\end{align}
For $N = 1$, from Eq.~\ref{eq:conn_part_one_ph} this correlation function is identical to the total transmission $\sum_{k=-\infty}^\infty |S_k(\nu)|^2$ through the localized system. For $N\geq 2$, this correlation function can be measured with $N-$photon coincidence counts on the emission from the localized system.

We now consider the calculation of a perturbative expansion for $G_N(\nu)$ with respect to $\Omega$. As is shown in appendix \ref{app:geometry}, it follows from the definition of the $N-$excitation Green's function that
\begin{subequations}\label{eq:pert_exp}
\begin{align}
\frac{1}{N!}\int_{-\infty}^\infty \dots \int_{-\infty}^\infty G(t, t \dots t; s_1, s_2 \dots s_N) \prod_{i=1}^N e^{-\textrm{i}\nu s_i}ds_i = e^{-\textrm{i}N\nu t}\bigg[\mathcal{G}_N^{(0)}(\text{p} (t); \nu) + \mathcal{G}_N^{(1)}(\text{p}(t); \nu)\cdot\frac{d\text{p}(t)}{dt} + O(\Omega^2)\bigg],
\end{align}
where $\mathcal{G}_N^{(0)}(\text{p}; \nu)$ is zeroth order in the modulation frequency $\Omega$ and is given by:
\begin{align}
\mathcal{G}_N^{(0)}(\text{p}; \nu) &= (-\textrm{i})^N\bra{g} L^N \bigg[\prod_{n=N}^1 (H_\text{eff}^n(\text{p}) - n\nu)^{-1}L^\dagger\bigg] \ket{g}
\end{align}
and $\mathcal{G}_N^{(1)}(\text{p}; \nu)$, also zeroth order in $\Omega$, is given by:
\begin{align}
\mathcal{G}_N^{(1)}(\text{p}; \nu) &= (-\textrm{i})^{N-1}\sum_{k=1}^N \bra{g}L^N\bigg[ \prod_{n=N}^{k+1}(H_\text{eff}^n(\text{p}) - n\nu)^{-1}L^\dagger\bigg](H_\text{eff}(\text{p}) - k\nu)^{-1}\nabla_\text{p}\bigg[\prod_{n=k}^1 (H_\text{eff}^n(\text{p}) - n\nu)^{-1} L^\dagger\bigg] \ket{g}.
\end{align}
\end{subequations}
Here $H_\text{eff}^n(\text{p}) $ is the $n-$excitation effective Hamiltonian as a function of the parameters $\text{p}$. The equal-time $N-$photon correlation function can now be expanded into a perturbative series in $\Omega$: $G_N(\nu) = G_N^{(0)}(\nu) + \Omega G_N^{(1)}(\nu) + O(\Omega^2)$ where both $G_N^{(0)}(\nu)$ and $G_N^{(1)}(\nu)$ are zeroth order in $\Omega$. It follows from Eqs.~\ref{eq:nph_corr_gfunc} and \ref{eq:pert_exp} that the zeroth order contribution $G_N^{(0)}(\nu)$ is given by:
\begin{align}
G_N^{(0)}(\nu) =  \int_0^T\big|\mathcal{G}_N^{(0)}(\text{p}(t); \nu)\big|^2\frac{dt}{T}.
\end{align}
It can be noted that $|\mathcal{G}_N^{(0)}(\text{p}; \nu)|^2$ is the equal-time $N-$photon correlation function that would be measured from the emission of a time-independent localized system with Hamiltonian $H_s(\text{p})$ and consequently to zeroth order $G_N(\nu)$ is simply a time-average of the instantaneous correlation function $|\mathcal{G}_N^{(0)}(\text{p}; \nu)|^2$. Furthermore, $G_N^{(0)}(\nu)$ is dynamical in nature \ie~it is dependent on the precise modulation of the parameters $\text{p}$. The first order contribution, $G_N^{(1)}(\nu)$, is given by:
\begin{align}
G_N^{(1)}(\nu) = \frac{1}{\pi} \text{Re}\bigg[\int_0^T \big[\mathcal{G}_N^{(0)}(\text{p}(t); \nu)\big]^* \mathcal{G}_N^{(1)}(\text{p}(t); \nu)\cdot \frac{d\text{p}(t)}{dt} dt \bigg] = \frac{1}{\pi} \text{Re}\bigg[\oint_\mathcal{C} \big(\mathcal{G}_N^{(0)}(\text{p}; \nu)\big)^* \mathcal{G}_N^{(1)}(\text{p}; \nu) \cdot d\text{p}\bigg].
\end{align}
It can immediately be seen that the first order correction $G_N^{(1)}(\nu)$ is completely geometric in nature \ie~it only depends on the loop $\mathcal{C}$ in the parameter space that the parameters $\text{p}$ trace during modulation.

\begin{figure}[t]
\centering
\includegraphics[scale=0.3]{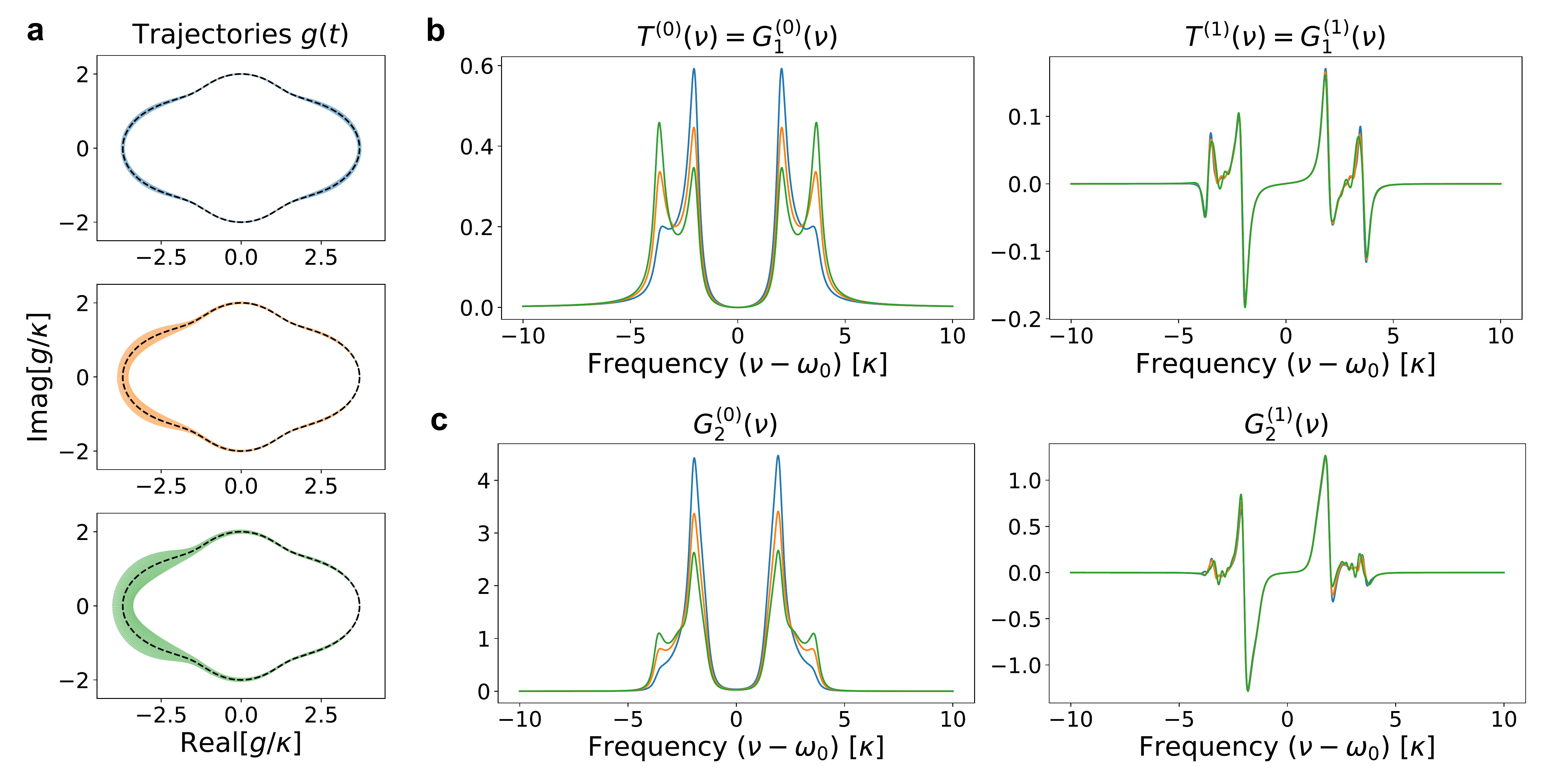}
\caption{\textbf{Scattering from a slowly modulated Jaynes Cumming system:} \textbf{a.} Modulation trajectories considered in our calculations --- the dashed line indicates the loop in the complex plane along which the cavity-TLS coupling constant $g$ is varied in one modulation period. The thickness of the colored shaded region around any point on the loop indicates how fast $g$ is changing at that point. \textbf{b.} Zeroth order and first order contribution to the total single-photon transmission $T(\nu) = G_1(\nu)$ through the Jaynes Cumming system as a function of the input frequency $\nu$. \textbf{c.} Zeroth order and first order contribution to the equal-time two photon correlation $G_2(\nu)$ in the output of the Jaynes Cumming system as a function of the input frequency $\nu$. }
\label{fig:slow_mod}
\end{figure}
As an illustrative example, we consider scattering from a Jaynes Cumming system formed by coupling a cavity with resonant frequency $\omega_c$ to a two-level system at frequency $\omega_e$:
\begin{align}
H_s(g) = \omega_e \sigma^\dagger \sigma + \omega_c a^\dagger a + (g a\sigma^\dagger + g^* a^\dagger \sigma),
\end{align}
where we modulate the complex cavity-TLS coupling strength $g$ periodically as a function of time to obtain a time-dependent Hamiltonian. We assume that this system coupled to the bosonic bath with through the cavity mode \ie~$L = \sqrt{\kappa / 2}\ a$ where $\kappa$ is the decay rate of the cavity. We consider three different modulations of $g$ as depicted in Fig.~\ref{fig:slow_mod}\text{a} which traverse the same loop in the complex plane per period. The shaded regions in Fig.~\ref{fig:slow_mod}a indicate the rate of change of $g$ with time at different points on the loop. Figure \ref{fig:slow_mod}b shows the zeroth and first order contributions to the transmission spectrum $T(\nu) = G_1(\nu)$ for the three different choices of $g(t)$. We clearly see that the zeroth order contribution $T^{(0)}(\nu)$ is dependent on the time-dependence of the modulation applied on $g$ whereas the first order contribution $T^{(1)}(\nu)$ is identical for the three different modulation schemes \ie~it is completely geometric in nature. A similar behavior can be seen for the zeroth and first order contributions to the equal-time two-photon correlation.

\section{Conclusion}
In this paper, we studied scattering of photons from periodically modulated quantum systems. A procedure for constructing $N-$photon scattering matrices and relating them to the Floquet decomposition of the effective Hamiltonian of the quantum system was outlined. Furthermore, we studied the properties of the equal time $N-$photon correlation function in the slow modulation regime and show that the first order correction with respect to the modulation frequency is completely geometric in nature. The formalism and results presented in this paper are of fundamental interest in the study of time-dependent open systems as well as for simulating quantum systems relevant for building quantum information processing systems.

\begin{acknowledgements}
The authors thank Momchil Minkov, Avik Dutt, Kevin Fischer, Daniil Lukin and Melissa Guidry for useful discussion. RT acknowledges support from Kailath Graduate Fellowship. ADW acknowledges support from Herb and Jane Dwight Graduate Fellowship. This research is funded by the U.S. Department of Energy, Office of Science, under Awards DE-SC0019174 and DE-Ac02-76SF00515, the National Science Foundation under award 1839056 and U.S.~Air Force Office of Scientific Research MURI project (Grant No.~FA9550-17-1-0002).
\end{acknowledgements}
\appendix
\section{Photon number conservation by the scattering matrix}\label{app:ph_no_cons}
Let $\Pi_s^n$ be the projector onto the $n^\text{th}$ excitation subspace $\mathcal{H}_s^n$. The excitation number operator $\mu_s$ can be constructed from $\Pi_s^n$ via:
\begin{align}
\mu_s = \sum_{n=0}^\infty n\Pi_s^n.
\end{align}
By construction, $\mu_s = \mu_s^\dagger$ and $\mu_s \ket{\phi} = n\ket{\phi}$ for $\ket{\phi} \in \mathcal{H}_s^n$. Additionally
 \begin{align}\label{eq:comm_l_mu}
 [L, \mu_s] = L. 
 \end{align}
 To see this, suppose $\ket{\phi} \in \mathcal{H}_s^n$ for any $n \geq 1$ then $L\ket{\phi} \in \mathcal{H}_s^{n-1}$. Therefore,
\begin{align}
L \mu_s \ket{\phi} = n L\ket{\phi} \text{ and } \mu_s L \ket{\phi} = (n - 1)L \ket{\phi} \implies [L, \mu_s]\ket{\phi} = L\ket{\phi}.
\end{align}
Furthermore, for $\ket{\phi} \in \mathcal{H}_s^0$, since $\mu_s \ket{\phi} = 0$ and $L\ket{\phi} = 0$ it follows that $[L, \mu_s] \ket{\phi} = 0 = L\ket{\phi}$. This shows that the operators $L$ and $\mu_s$ satisfy Eq.~\ref{eq:comm_l_mu}

Finally, consider the excitation number operator $\mu$ for the full system constructed by adding $\mu_s$ with the photon number operator for the two baths,
\begin{align}\label{eq:ex_num_op}
\mu = \mu_s + \sum_{l \in \{a, b\}} \int_{-\infty}^\infty l_\omega^\dagger l_\omega d\omega.
\end{align}
From the commutator $[L, \mu_s] = L$ it follows that $[H(t), \mu] = 0$ \ie~the observable corresponding to $\mu$ is a conserved quantity. Since a state with $N$ photons in the bosonic baths and the system in $\ket{g}$ is an eigenstate of $\mu$ with eigenvalue $N$, this conservation law immediately implies that it can only scatter into a state with $N$ photons in the bosonic baths.

\section{Relating the scattering matrix elements to the Green's function}\label{app:smat_to_gfunc}
In this appendix, we derive the relationship between the Green's function $G(t_1, t_2 \dots t_N; s_1, s_2 \dots s_N)$ and the scattering matrix element $S(\omega_1, \omega_2 \dots \omega_N; \nu_1, \nu_2 \dots \nu_N)$ (Eq.~\ref{eq:gfunc_to_smat}). Using the fact that $H_s(t)\ket{g} = 0$ and $L\ket{g} = 0$, it follows from Eq.~\ref{eq:basic_hamiltonian} that $H(t)\ket{g; \text{vac}} = 0$. Noting that the propagator $U_0(t_f, t_i)$ with respect to the Hamiltonian $H_0(t)$ (Eq.~\ref{eq:ref_hamil}) satisfies $U_0(t_i, t_f) l_\omega U_0(t_f, t_i) = l_\omega e^{-\textrm{i}\omega (t_f - t_i)} \ \forall \ l \in\{a, b\}$, the scattering matrix element $S(\omega_1, \omega_2 \dots \omega_N; \nu_1, \nu_2 \dots \nu_N)$ can be expressed as:
\begin{align}\label{eq:smat_in_terms_hop}
S(\omega_1, \omega_2 \dots \omega_N; \nu_1, \nu_2 \dots \nu_N) = e^{-\textrm{i}\phi(t_0)} \lim_{\substack{t_+\to \infty \\ t_- \to -\infty}}e^{\textrm{i}\sum_{i=1}^N (\omega_i t_+ - \nu_i t_-)} \bra{g;\text{vac}} \mathcal{T}\bigg[\bigg(\prod_{i=1}^N b_{\omega_i}(t_+)\bigg) \bigg(\prod_{i=1}^N a_{\nu_i}^\dagger(t_-)\bigg)\bigg]\ket{g;\text{vac}},
\end{align}
where $a_\nu(t_-) = U(t_0, t_-) a_\nu U(t_-, t_0)$ and $b_\omega(t_+) = U(t_0, t_+) b_\omega U(t_+, t_0)$ with $U(\cdot, \cdot)$ being the propagator with respect to the Hamiltonian $H(t)$ (Eq.~\ref{eq:basic_hamiltonian}) and $\mathcal{T}[\cdot]$ indicates a time-ordering with respect to its arguments. Note that since $t_+ \geq t_-$, this time-ordering is effectively an identity operation in Eq.~\ref{eq:smat_in_terms_hop}. Next, we use the Heisenberg equations of motion for $a_\nu(t)$ and $b_\omega(t)$ --- from Eq.~\ref{eq:basic_hamiltonian}, it follows that:
\begin{align}
\frac{d}{dt} \begin{pmatrix}
a_\nu(t) \\
b_\omega(t)
\end{pmatrix} = -\textrm{i}\begin{pmatrix}
\nu a_\nu(t) \\
\omega b_\omega(t)
\end{pmatrix} + \frac{1}{\sqrt{2\pi}} \begin{pmatrix}
L(t) \\
L(t)
\end{pmatrix}
\end{align}
These equations of motion can easily be integrated from $t_-$ to $t_+$ to yield the following:
\begin{subequations}
\begin{align}
&a_\nu^\dagger(t_-) = a_\nu^\dagger(t_+)e^{-\textrm{i}\nu(t_+ - t_-)} - \int_{t_-}^{t_+} L(s) e^{-\textrm{i}\nu (s - t_-)}\frac{ds}{\sqrt{2\pi}}\label{eq:a_init_final}\\
&b_\omega(t_+) = b_\omega(t_-)e^{-\textrm{i}\omega(t_+ - t_-)} + \int_{t_-}^{t_+} L(t) e^{-\textrm{i}\omega( t_+ - t)}\frac{dt}{\sqrt{2\pi}}\label{eq:b_init_final}
\end{align}
\end{subequations}
Substituting Eq.~\ref{eq:b_init_final} into Eq.~\ref{eq:smat_in_terms_hop} and noting that any term with $b_{\omega_i}(t_-)$ goes to 0 since the time-ordering operator places it to the right of $L(t)\ \forall \ t \in (t_-, t_+)$, $b_{\omega_i}(t_-)$ commutes with $a_{\nu_i}(t_-)$ and annihilates $\ket{g; \text{vac}}$, we obtain:
\begin{align}\label{eq:smat_partial}
&S(\omega_1, \omega_2 \dots \omega_N; \nu_1, \nu_2 \dots \nu_N)\nonumber\\
&= e^{-\textrm{i}\phi(t_0)} \lim_{\substack{t_+\to\infty \\ t_- \to -\infty}} \int_{t_-}^{t_+}\dots \int_{t_-}^{t_+} \bra{g; \text{vac}} \mathcal{T}\bigg[\bigg(\prod_{i=1}^N L(t_i)\bigg)\bigg(\prod_{i=1}^N a_{\nu_i}^\dagger(t_-)\bigg) \bigg]\ket{g;\text{vac}}\prod_{i=1}^N e^{\textrm{i}\omega_i t_i} \frac{dt_i}{\sqrt{2\pi}}
\end{align}
Similarly, substituting Eq.~\ref{eq:a_init_final} into Eq.~\ref{eq:smat_partial} and noting that any term with $a_{\nu_i}(t_+)$ goes to 0 since the time-ordering operator places it to the left of $L(t), L(t) \ \forall \ t \in(t_-, t_+)$ and that $a^\dagger_{\nu_i}(t_+)$ annihilates $\bra{g; \text{vac}}$, we obtain the result in Eq.~\ref{eq:gfunc_to_smat}.

\section{Scattering matrix calculation}\label{app:smat}
\subsection{Single-photon scattering matrix}
The starting point for the calculation of the single-photon scattering matrix is the evaluation of the single-photon Green's function which is given by:
\begin{align}\label{eq:single_ex_gfunc}
G(t; s) = \bra{ g} \mathcal{T}[\tilde{L}(t) \tilde{L^\dagger}(s)]\ket{g} = \bra{g} L {U}_\text{eff}^1(t, s)L^\dagger\ket{g}\Theta(t\geq s).
\end{align}
Using the Floquet decomposition of $U_\text{eff}^1(t, s)$ (Eq.~\ref{eq:floquet_decomp}), this can be expressed as:
\begin{align}
G(t; s) =  \big(\text{L}^{1\to 0}(t) \text{D}\big[e^{-\textrm{i}\lambda^1(t - s)}\big]  \text{L}^{0\to 1}(s)\big)\Theta(t\geq s),
\end{align}
where $\text{L}^{1\to0}(t)$ is a row-vector and $L^{0 \to 1}(s)$ is a column-vector, and their elements are given by:
\begin{align}\label{eq:td_ann_mat_elm}
\big[\text{L}^{1\to 0}(t)\big]_n = \bra{g} L \ket{\phi_n(t)} \ \text{ and } \ \big[\text{L}^{0 \to 1}(s)\big] = \bra{\chi_n^1(s)} L^\dagger \ket{g}
\end{align} 
We note that $\text{L}^{1\to 0}_k$ and $\text{L}^{0\to 1}_k$ defined in Eq.~\ref{eq:L_fc_def_single_ex} of the main text are simply the Fourier series coefficients of $\text{L}^{1 \to 0}(t)$ and $\text{L}^{0\to 1}(s)$ respectively:
\begin{align}\label{eq:fseries}
\text{L}^{1\to 0}(t) = \sum_{k \in \mathbb{Z}} \text{L}^{1\to 0}_k e^{-\textrm{i}k\Omega t} \ \text{and} \ \text{L}^{0\to 1}(s) = \sum_{k\in \mathbb{Z}} \text{L}^{0\to 1}_k e^{\textrm{i}k\Omega s}.
\end{align}
From Eqs.~\ref{eq:gfunc_to_smat}, \ref{eq:single_ex_gfunc} and \ref{eq:fseries}, it follows that the single-photon scattering matrix $S(\omega; \nu)$ is given by Eq.~\ref{eq:conn_part_one_ph} in the main text.

\subsection{Two-photon scattering matrix} \label{app:two_ph_smat}
The two-excitation Green's function $G(t_1, t_2; s_1, s_2)$, given by Eq.~\ref{eq:gfunc_eff_hamil} with $N = 2$, is symmetric under the swap operations $t_1\leftrightarrow t_2$ and $s_1 \leftrightarrow s_2$. Defining $\mathcal{G}(t_1, t_2; s_1, s_2) = G(t_1, t_2; s_1, s_2) \Theta(t_1\geq t_2 \ \text{and} \ s_1\geq s_2)$, it follows that:
\begin{align}\label{eq:ordered_gfunc}
G(t_1, t_2; s_1, s_2) = \sum_{P, Q} \mathcal{G}(t_{P(1)}, t_{P(2)}; s_{Q(1)}, s_{Q(2)}),
\end{align}
where $P, Q$ are permutations of the two-element set $\{1, 2\}$. It thus follows from Eq.~\ref{eq:gfunc_to_smat} that:
\begin{align}\label{eq:full_smat_partial_smat}
S(\omega_1, \omega_2; \nu_1, \nu_2) = \sum_{P, Q}e^{-\textrm{i}\phi(t_0)} \mathcal{S}(\omega_{P(1)}, \omega_{P(2)}; \nu_{Q(1)}, \nu_{Q(2)}),
\end{align}
where
\begin{align}\label{eq:def_new_S}
\mathcal{S}(\omega_1, \omega_2; \nu_1, \nu_2) = \int_{-\infty}^\infty \dots \int_{-\infty}^\infty \mathcal{G}(t_1, t_2; s_1, s_2) \prod_{i=1}^2 e^{\textrm{i}(\omega_i t_i  - \nu_i s_i)}\frac{dt_i ds_i}{2\pi}.
\end{align}
From Eq.~\ref{eq:gfunc_eff_hamil}, it follows that:
\begin{align}
\mathcal{G}(t_1, t_2; s_1, s_2) = \mathcal{G}^1(t_1, t_2; s_1, s_2) + \mathcal{G}^2(t_1, t_2; s_1, s_2),
\end{align}
where
\begin{subequations}
\begin{align}
\mathcal{G}^1(t_1, t_2; s_1, s_2) &= \bra{g} L(t_1) L^\dagger(s_1) L(t_2) L^\dagger(s_2) \ket{g} \Theta(t_1\geq s_1 \geq t_2 \geq s_2) \nonumber \\
& =\bra{g}L U_\text{eff}^1(t_1, s_1) L^\dagger \ket{g}\bra{g}L U_\text{eff}^1(t_2, s_2) L^\dagger \ket{g}\Theta(t_1\geq s_1 \geq t_2 \geq s_2)
 \\ \mathcal{G}^2(t_1, t_2; s_1, s_2) &= \bra{g}L(t_1) L(t_2) L^\dagger(s_1) L^\dagger(s_2)\ket{g} \Theta(t_1\geq t_2 \geq s_1 \geq s_2) \\
 &=\bra{g}L U_\text{eff}^1(t_1, t_2) L^\dagger U_\text{eff}^2(t_2, s_1) L U_\text{eff}^1(s_1, s_2)L^\dagger\ket{g}\Theta(t_1\geq t_2 \geq s_1 \geq s_2).
\end{align}
\end{subequations}
Using the Floquet decomposition of $U_\text{eff}^{1,2}(t, s)$ (Eq.~\ref{eq:floquet_decomp}), it follows that:
\begin{subequations}\label{eq:g2_explicit}
\begin{align}
\mathcal{G}^1&(t_1, t_2; s_1, s_2) =\nonumber\\ &\big(\text{L}^{1 \to 0}(t_1) \text{D}\big[e^{-\textrm{i}\lambda^1(t_1 - s_1)}\big] \text{L}^{0 \to 1}(s_1)\big)\big(\text{L}^{1 \to 0}(t_2) \text{D}\big[e^{-\textrm{i}\lambda^1(t_2 - s_2)}\big] \text{L}^{0 \to 1}(s_2)\big) \Theta(t_1\geq s_1 \geq t_2 \geq s_2), \\
\mathcal{G}^2&(t_1, t_2; s_1, s_2) = \nonumber \\ &\big(\text{L}^{1\to 0}(t_1) \text{D}\big[e^{-\textrm{i}\lambda^1(t_1 - t_2)}\big] \text{L}^{2\to 1}(t_2) \text{D}\big[e^{-\textrm{i}\lambda^2(t_2 - s_1)}\big] \text{L}^{1\to 2}(s_1) \text{D}\big[e^{-\textrm{i}\lambda^1(s_1 - s_2)}\big] \text{L}^{1\to 0}(s_2) \big) \Theta(t_1 \geq t_2 \geq s_1 \geq s_2),
\end{align}
where $\text{L}^{1\to 0}(t), \text{L}^{0\to 1}(s)$ are defined in Eq.~\ref{eq:td_ann_mat_elm} and $L^{2\to 1}(t), L^{1\to 2}(s)$ are matrices with elements 
\begin{align}
\big[\text{L}^{2\to 1}(t)\big]_{m, n} = \bra{\chi^1_m(t)} L \ket{\phi^2_n(t)} \ \text{and} \ \big[\text{L}^{1\to 2}(s)\big]_{m, n} = \bra{\chi_m^2(s)}L^\dagger \ket{\phi_n^1(s)}.
\end{align}
\end{subequations}
It can be noted that $\text{L}^{2\to1}_k$ and $\text{L}^{1\to 2}_k$ defined in Eq.~\ref{eq:L_fc_def_two_ex} of the main text are simply the Fourier series coefficients of $\text{L}^{2\to 1}(t)$ and $\text{L}^{1\to 2}(s)$:
\begin{align}
\text{L}^{2\to 1}(t) = \sum_{k \in \mathbb{Z}} L_k^{2\to 1} e^{-\textrm{i}k\Omega t} \ \text{and} \ \text{L}^{1\to 2}(s) = \sum_{k\in \mathbb{Z}} L_k^{1\to 2} e^{\textrm{i}k\Omega s}.
\end{align}
Using Eq.~\ref{eq:g2_explicit} to evaluate the integral in Eq.~\ref{eq:def_new_S}, we obtain:
\begin{subequations}\label{eq:partial_smat}
\begin{align}
\mathcal{S}(\omega_1, \omega_2; \nu_1, \nu_2) = \sum_{\substack{k\in \mathbb{Z}, \\ j \in\{1,2\}}} \mathcal{S}^j_k(\omega_1, \omega_2; \nu_1, \nu_2)  \delta(\omega_1 + \omega_2 - \nu_1 - \nu_2 - k\Omega),
\end{align}
where
\begin{alignat}{2}
\mathcal{S}^{1}_k(\omega_1, \omega_2; \nu_1, \nu_2) &= \frac{1}{2\pi \textrm{i}}\sum_{p, m, n \in \mathbb{Z}} \bigg[&&\text{L}^{1\to 0}_p \text{D}\bigg(\frac{1}{\textrm{i}(\uplambda^1 - \omega_1 + p\Omega)} \bigg)\text{L}^{0\to 1}_{n + p - k} \bigg(\frac{1}{\textrm{i}(\omega_2 - \nu_2-n\Omega) - \textrm{i} 0^+}\bigg) \nonumber \\& &&\text{L}^{1\to 0}_{m + n} \text{D}\bigg(\frac{1}{\textrm{i}(\lambda^1 - \nu_2 + m\Omega) } \bigg)\text{L}^{0\to 1}_{m}\bigg],\\
\mathcal{S}^{2}_k(\omega_1, \omega_2; \nu_1, \nu_2) &= \frac{1}{2\pi}\sum_{p, m, n \in \mathbb{Z}} \bigg[&&\text{L}^{1\to 0}_p \text{D}\bigg(\frac{1}{\textrm{i}(\lambda^1 - \omega_1 + p\Omega)} \bigg)\text{L}^{2\to 1}_{n - p + k} \text{D}\bigg(\frac{1}{\textrm{i}(\lambda^2 - \nu_1 - \nu_2 + n\Omega) } \bigg) \nonumber \\& &&\text{L}^{1\to 2}_{n - m} \text{D}\bigg(\frac{1}{\textrm{i}(\lambda^1 - \nu_2 + m\Omega)} \bigg)\text{L}^{0\to 1}_{m}\bigg].
\end{alignat}
\end{subequations}
The full two-photon scattering matrix can be constructed from Eqs.~\ref{eq:full_smat_partial_smat} and \ref{eq:partial_smat}. To explicitly extract the connected part of the two-photon scattering matrix, we note that $1 / (x - \textrm{i}0^+) = \mathcal{P}(1/x) + \pi \delta(x)$ --- the connected part of the scattering matrix can thus be obtained by replacing $1 / (x - \textrm{i}0^+)$ by $\mathcal{P}(1/x)$ in the resulting expressions for the scattering matrix. This yields the results in Eq.~\ref{eq:conn_part_two_ph} in the main text.

Finally, we show that $S^{\text{C},1}_k(\omega_1, \omega_2; \nu_1, \nu_2)$ defined in Eq.~\ref{eq:conn_part_two_ph} is not singular despite containing the principle parts. We begin by rewriting it as
\begin{alignat}{2}
S^{\text{C}, 1}_k(\omega_1, \omega_2;& \nu_1, \nu_2) = \sum_{P, Q} \sum_{n\in \mathbb{Z}} S_{k - n}(\omega_{P(1)} - (k - n)\Omega) S_n(\nu_{Q(2)}) \mathcal{P}\frac{1}{\omega_{P(2)} - \nu_{Q(2)} - n\Omega}, \nonumber \\
=&\sum_{n\in \mathbb{Z}}\sum_{i=1}^2 \bigg[ S_{k - n}(\omega_i - (k - n)\Omega) S_n(\bar{\nu}_i) \mathcal{P}\frac{1}{\bar{\omega}_i - \bar{\nu}_i - n\Omega} + S_{k- n }(\bar{\omega}_i - (k - n)\Omega) S_n({\nu}_i) \mathcal{P}\frac{1}{{\omega}_i - {\nu}_i - n\Omega}\bigg] \nonumber \\
=&\sum_{n\in \mathbb{Z}} \sum_{i=1}^2 \bigg[ S_{n}(\omega_i - n\Omega) S_{k-n}(\bar{\nu}_i) \mathcal{P}\frac{1}{\bar{\omega}_i - \bar{\nu}_i - (k - n)\Omega} + S_{k - n}(\bar{\omega}_i - (k - n)\Omega) S_{n}({\nu}_i) \mathcal{P}\frac{1}{{\omega}_i - {\nu}_i - n\Omega}\bigg]
\end{alignat}
where $\bar{\lambda}_{1, 2} = \lambda_{2, 1}$ for $\lambda \in \{\omega, \nu\}$ and $S_k(\nu)$ is the scattering amplitude of a single-photon at frequency $\nu$ into frequency $\nu + k\Omega$ defined in Eq.~\ref{eq:conn_part_one_ph}. We note that since the arguments of $S^{\text{C},1}_k$ are constrained to satisfy $\omega_1 + \omega_2 - \nu_1 - \nu_2 = k\Omega$, it follows that:
\begin{align}
S_k^{\text{C}, 1}(\omega_1, \omega_2; \nu_1, \nu_2) = \sum_{n\in \mathbb{Z}} \sum_{i=1}^2 \bigg[S_{n}(\nu_i - (\nu_i - \omega_i + n\Omega))S_{k-n}(\bar{\nu}_i)- S_{k-n}(\bar{\nu}_i + (\nu_i - \omega_i + n\Omega))S_n(\nu_i)\bigg]\mathcal{P}\frac{1}{\nu_i - \omega_i + n\Omega}
\end{align}
Noting that
\begin{align}
S_n(\nu - \delta) S_{m}(\bar{\nu}) - S_n(\nu)  S_{m}(\bar{\nu} + \delta) = M_{n, m}(\nu, \bar{\nu}, \delta) \delta
\end{align}
where $M_{n, m}(\nu, \bar{\nu}, \delta)$, defined below, is a smooth and finite function of its arguments:
\begin{align}
M_{n, m}(\nu, \bar{\nu}, \delta) = \sum_{i, j}\sum_{p, q \in \mathbb{Z}}\frac{\big[\text{L}_{p}^{0\to 1} \big]_i \big[\text{L}_{p + n}^{1\to 0}\big]_i\big[\text{L}_q^{0\to1}\big]_j\big[\text{L}^{1\to0}_{q+m}\big]_j (\lambda_j^1 + \lambda_i^1 + (p + q)\Omega - \bar{\nu} - \nu)}{(\lambda_i^1 + p\Omega - (\nu - \delta))(\lambda_j^1 + q\Omega - \bar{\nu})(\lambda_i^1 + p\Omega - \textrm{i}\nu)(\lambda_j^1 + q\Omega - (\bar{\nu} + \delta))},
\end{align}
it follows that
\begin{align}
S_k^{\text{C}, 1}(\omega_1, \omega_2; \nu_1, \nu_2) = \sum_{n\in \mathbb{Z}}\sum_{i= 1}^2 M_{n, k-n}(\nu_i, \bar{\nu}_i; \nu_i - \omega_i + n\Omega).
\end{align}
This shows that $S_k^{\text{C}, 1}(\omega_1, \omega_2; \nu_1, \nu_2)$ is indeed a well defined and finite function of the input and output frequencies subject to the constraint $\omega_1 + \omega_2 - \nu_1 - \nu_2 = k\Omega$. 
\section{Equal time $N$-photon correlation function in slow modulation regime}\label{app:geometry}
In this appendix, we provide a derivation of Eq.~\ref{eq:pert_exp} of main text. Noting that the $N-$excitation Green's function $G(t_1, t_2 \dots t_N; s_1, s_2\dots s_N)$ is symmetric under permutation of the times $s_1, s_2 \dots s_N$, it follows that:
\begin{align}\label{eq:g_ordered}
\frac{1}{N!} \int_{-\infty}^\infty \dots \int_{-\infty}^\infty G(t, t \dots t; s_1, s_2 \dots s_N) \prod_{i=1}^N e^{-\textrm{i}\nu s_i} ds_i = \int_{-\infty}^{\infty} \dots \int_{-\infty}^{s_3}\int_{-\infty}^{s_2} G(t, t \dots t; s_1, s_2 \dots s_N) \prod_{i=1}^N e^{-\textrm{i}\nu s_i} ds_i.
\end{align}
Furthermore, from Eq.~\ref{eq:gfunc_eff_hamil}, we obtain that if $s_1 \leq s_2 \dots \leq s_N$ then:
\begin{align}
G(t, t \dots t; s_1, s_2 \dots s_N) = \langle g | L^N \bigg[\prod_{n = {N}}^1 U_\text{eff}^n(s_{n + 1}, s_n) L^\dagger\bigg]_{s_{N+1} = t} \ket{g}
\end{align}
To proceed further, we consider the evaluation of
\begin{align}
 \int_{-\infty}^t U_\text{eff}^n(t, s) O(s) e^{-\textrm{i} \nu s} ds,
\end{align}
where $O(s)$ is a time-dependent operator which is assumed to be slowly varying. Since $U_\text{eff}^n(t, s)$ is the propagator corresponding to the Hamiltonian $H_\text{eff}^n(t)$, it follows that:
\begin{align}
\int_{-\infty}^t U_\text{eff}^n(t, s) O(s) e^{-\textrm{i} \nu s} ds = -\textrm{i}\int_{-\infty}^t \bigg[\frac{\partial}{\partial s}\big(U_\text{eff}^n(t, s) e^{-\textrm{i} \nu s}\big)\bigg] (H_\text{eff}^n(s) - \nu)^{-1}O(s) ds.
\end{align}
Applying integration by parts, we obtain:
\begin{align}\label{eq:one_by_parts}
\int_{-\infty}^t U_\text{eff}^n(t, s) O(s) e^{-\textrm{i} \nu s} ds= -\textrm{i}(H_\text{eff}^n(t) - \nu)^{-1}O(t)e^{-\textrm{i}\nu t} + \textrm{i}\int_{-\infty}^t U_\text{eff}^n(t, s) \frac{\partial}{\partial s}\big[ (H_\text{eff}^n(s) - \nu)^{-1}O(s)\big] e^{-\textrm{i}\nu s}ds.
\end{align}
Repeating a similar calculation for the integral on the right of Eq.~\ref{eq:one_by_parts} and neglecting terms that are second order in the derivatives of $H_\text{eff}^n(t)$ and $O(t)$, we obtain:
\begin{align}
\int_{-\infty}^t U_\text{eff}^n(t, s) O(s) e^{-\textrm{i} \nu s} ds \approx -\textrm{i}(H_\text{eff}^n(t) - \nu)^{-1}O(t) e^{-\textrm{i}\nu t} + (H_\text{eff}^n(t) - \nu)^{-1} \frac{\partial}{\partial t} \big[ (H_\text{eff}^n(s) - \nu)^{-1}O(s)\big]  e^{-\textrm{i}\nu t}
\end{align}
Repeated application of this to the integral in Eq.~\ref{eq:g_ordered} together with neglecting any terms that second order or higher in the derivatives of effective Hamiltonian, we obtain the result in Eq.~\ref{eq:pert_exp}.

\bibliography{library}{}
\end{document}